\documentclass[twocolumn, times, tighten]{aastex63}
\usepackage{textcomp}
\usepackage{amsmath}
\usepackage{color}
\usepackage{xcolor}
\usepackage{amssymb}
\usepackage{hyperref}

\newcommand{\MgII}{Mg\,{\sc ii}}
\newcommand{\CIV}{C\,{\sc iv}}
\newcommand{\CIII}{C\,{\sc iii]}}

\submitjournal{ApJ}

\shortauthors{Sun et al.}
\shorttitle{CHAR}

\begin{document}
\nocite{*}

\title{Modeling Quasar UV/Optical Variability with the Corona-heated 
Accretion-disk Reprocessing (CHAR) Model}

\author[0000-0002-0771-2153]{Mouyuan Sun}
\affiliation{Department of Astronomy, Xiamen University, Xiamen, Fujian 
361005, China; msun88@xmu.edu.cn}

\author[0000-0002-1935-8104]{Yongquan Xue}
\affiliation{CAS Key Laboratory for Research in Galaxies and Cosmology, 
Department of Astronomy, University of Science and Technology of China, 
Hefei 230026, China}
\affiliation{School of Astronomy and Space Science, University of Science 
and Technology of China, Hefei 230026, China}

\author[0000-0001-8416-7059]{Hengxiao Guo}
\affiliation{Department of Astronomy, University of Illinois at Urbana-Champaign, 
Urbana, IL 61801, USA}
\affiliation{National Center for Supercomputing Applications, University of Illinois 
at Urbana-Champaign, Urbana, IL 61801, USA}
\affiliation{Department of Physics and Astronomy, 4129 Frederick Reines Hall, University 
of California, Irvine, CA, 92697-4575, USA}

\author[0000-0002-4419-6434]{Junxian Wang}
\affiliation{CAS Key Laboratory for Research in Galaxies and Cosmology, 
Department of Astronomy, University of Science and Technology of China, 
Hefei 230026, China}
\affiliation{School of Astronomy and Space Science, University of Science 
and Technology of China, Hefei 230026, China}

\author[0000-0002-0167-2453]{W. N. Brandt}
\affiliation{Department of Astronomy \& Astrophysics, 525 Davey Lab, The 
Pennsylvania State University, University Park, PA 16802, USA}
\affiliation{Institute for Gravitation and the Cosmos, 525 Davey Lab, The 
Pennsylvania State University, University Park, PA 16802, USA}
\affiliation{Department of Physics, 104 Davey Lab, The Pennsylvania State 
University, University Park, PA 16802, USA}

\author[0000-0002-1410-0470]{Jonathan R. Trump}
\affiliation{Department of Physics, University of Connecticut, Storrs, CT 06269, 
USA}

\author[0000-0003-3667-1060]{Zhicheng He}
\affiliation{CAS Key Laboratory for Research in Galaxies and Cosmology, 
Department of Astronomy, University of Science and Technology of China, 
Hefei 230026, China}
\affiliation{School of Astronomy and Space Science, University of Science 
and Technology of China, Hefei 230026, China}

\author[0000-0001-8678-6291]{Tong Liu}
\affiliation{Department of Astronomy, Xiamen University, Xiamen, Fujian 
361005, China; msun88@xmu.edu.cn}

\author[0000-0001-7349-4695]{Jianfeng Wu}
\affiliation{Department of Astronomy, Xiamen University, Xiamen, Fujian 
361005, China; msun88@xmu.edu.cn}

\author{Haikun Li}
\affiliation{Department of Astronomy, Xiamen University, Xiamen, Fujian 
361005, China; msun88@xmu.edu.cn}

\begin{abstract}
The rest-frame UV/optical variability of the quasars in the Sloan Digital Sky 
Survey (SDSS) Stripe 82 is used to test the Corona-Heated Accretion-disk 
Reprocessing (CHAR) model of \cite{Sun2020}. We adopt our CHAR model and the 
observed black-hole masses ($M_{\mathrm{BH}}$) and luminosities ($L$) 
to generate mock light curves that share the same measurement noise and 
sampling as the real observations. Without any fine-tuning, 
our CHAR model can satisfactorily reproduce the observed ensemble structure 
functions for different $M_{\mathrm{BH}}$, $L$, and rest-frame 
wavelengths. Our analyses reveal that a luminosity-dependent bolometric correction 
is disfavored over the constant bolometric correction for UV/optical luminosities. 
Our work demonstrates the possibility of extracting quasar 
properties (e.g., the bolometric correction or the dimensionless viscosity parameter) 
by comparing the physical CHAR model with quasar light curves. 
\end{abstract}


\section{Introduction}
\label{sect:intro}
AGN UV/optical variability offers a new way to resolve the broad emission-line regions 
\citep{Blandford1982} as well as the accretion disks \citep{Collin1991, Krolik1991} 
and probe the density of the outflowing gas density \citep{He2019, Li2019}. AGN UV/optical 
variability is most likely to be driven by time-dependent evolution of the central 
engine (i.e., accretion disk) because many studies \citep{Kelly2009,Macleod2010, 
Sun2015,Caplar2017} found that AGN UV/optical variability depends at least 
on black-hole mass ($M_{\mathrm{BH}}$) and luminosity ($L$). However, 
our physical understanding of AGN UV/optical variability is far from clear 
\citep{Lawrence2018}. In a 
previous work \citep{Sun2020}, we proposed a Corona-Heated Accretion-disk Reprocessing 
(CHAR) model to explain the UV/optical variability of quasars.\footnote{We use the term 
quasars to generically refer to active galactic nuclei (AGNs) with optical broad emission 
lines, regardless of luminosity. That is, AGNs and quasars are used interchangeably in this 
work.} In the CHAR model, the X-ray corona and the underlying cold accretion disk are 
magnetically coupled. Coronal magnetic fluctuations can induce coherent fluctuations in the 
disk heating rate which alter the disk temperature and UV/optical luminosity. We demonstrated 
that the CHAR model can explain high-quality \textit{Kepler} AGN light curves, as well as the 
larger-than-expected inter-band time lags and the multiwavelength structure functions (i.e., 
the variability amplitude as a function of timescale; see also Section~\ref{sect:sample}) of 
NGC 5548. We also showed that the CHAR model has the potential to explain the dependence of 
AGN UV/optical variability parameters upon $M_{\mathrm{BH}}$, $L$, and rest-frame 
wavelength ($\lambda_{\mathrm{rest}}$). In \cite{Sun2020}, 
we additionally suggested that the CHAR model could be used to fit the observed UV/optical 
variability of quasars and laid out a plan for future work that would make a more detailed 
comparison between the CHAR model predictions and the observational results of the correlations 
between quasar UV/optical variability and physical properties. The Sloan Digital Sky Survey 
\citep[SDSS;][]{Gunn2006} Stripe 82 (hereafter SDSS S82) quasar observations provide a valuable 
dataset for fitting by our physical CHAR model. Compared to adopting empirical stochastic models 
(e.g., the popular CAR(1) model, a.k.a., the damped random walk model) to fit AGN light curves 
\citep[e.g.,][]{Kelly2009, Macleod2010}, our modeling results have straightforward physical 
implications. 

This paper is formatted as follows. In Section~\ref{sect:sample}, we introduce the SDSS S82 
quasar light curves and the corresponding ensemble structure functions. In Section~\ref{sect:mock}, 
we present our mock ensemble structure functions and compare them with the observed ones. 
In Section~\ref{sect:dis}, we discuss our results and present a recipe to simulate quasar 
multi-band stochastic light curves. Our conclusions are summarized in Section~\ref{sect:sum}. 
The Schwarzschild radius is $R_{\mathrm{S}}\equiv 2GM_{\mathrm{BH}}/c^2$, where $G$ and $c$ 
are the gravitational constant and speed of light, respectively.

\section{Sample Construction and Observed Variability}
\label{sect:sample}
\begin{figure*}
\epsscale{1.0}
\plotone{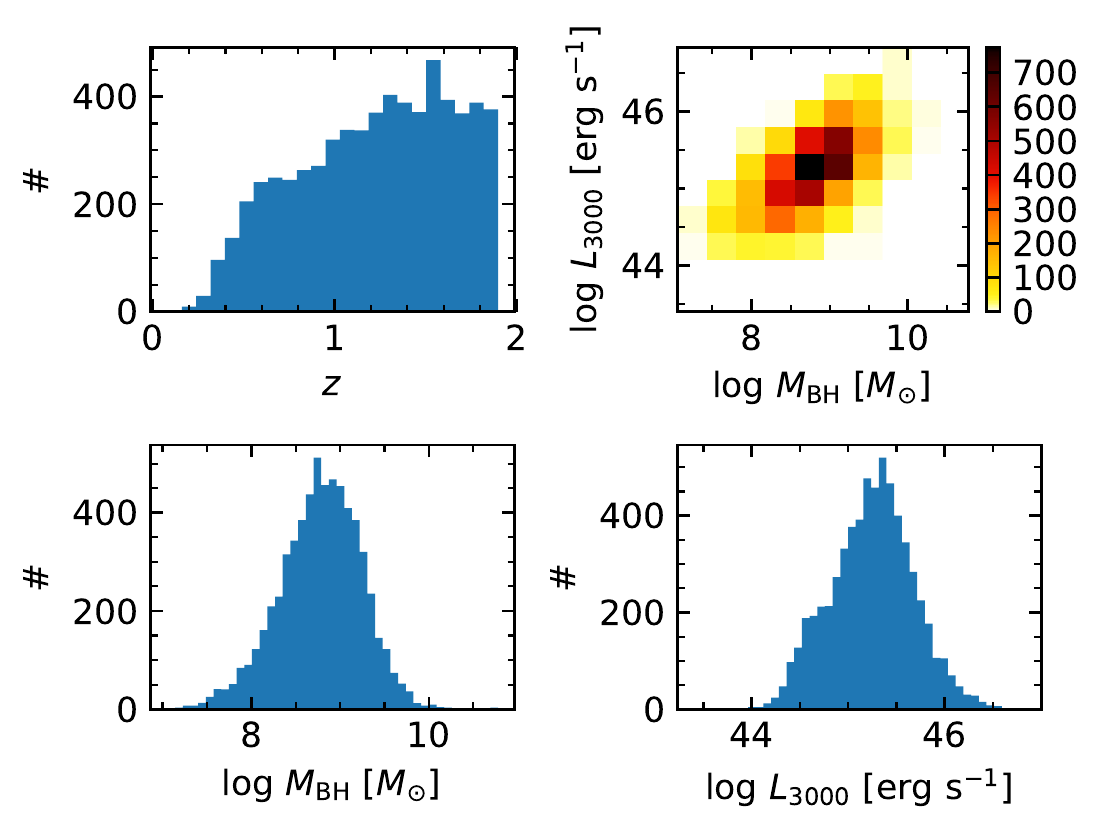}
\caption{Distributions of redshift, $M_{\mathrm{BH}}$, and $L_{\mathrm{bol}}$ for the parent 
sample of the SDSS S82 quasars. Most sources have $z>1$ (the upper-left panel) and span more 
than $2$ dex in $M_{\mathrm{BH}}$ and $L_{\mathrm{bol}}$. \label{fig:pms}}
\end{figure*}

Following \cite{Macleod2010}, we consider the variability data of SDSS S82 quasars with 
multi-epoch (on average $>60$ epochs) and multi-band \citep[i.e., $ugriz$; see][]{Fukugita1996} 
observations. The light curves\footnote{These data can be downloaded from 
\url{http://faculty.washington.edu/ivezic/macleod/qso_dr7/Southern.html}.} are well calibrated 
by \cite{Ivezic2007} and \cite{Sesar2007}. First, we cross-match these quasars with the catalog 
of SDSS DR7 quasar properties \citep{Shen2011} and obtain their redshifts ($z$), $M_{\mathrm{BH}}$, 
and the rest-frame $\lambda = 3000\ \mathrm{\AA}$ luminosity (i.e., $L_{\mathrm{3000}} = \lambda 
L_{\lambda}$ for the rest-frame $\lambda = 3000\ \mathrm{\AA}$). 
Second, we only select sources with either H$\beta$ or \MgII\ virial black-hole mass 
estimates (i.e., $z<1.9$); we do not adopt the \CIV\ black-hole mass estimator 
since \CIV\ often shows (non-virial) outflow signatures \citep[e.g.,][]{Richards2011, Denney2012, 
Sun2018b} and the scatter of the ratio of the \CIV-based $M_{\mathrm{BH}}$ to the H$\beta$-based 
one is substantial ($\lesssim 1$ dex) unless detailed empirical corrections are applied 
\citep[e.g.,][]{Coatman2017, Marziani2019, Zuo2020}. Radio-loud 
(i.e., radio loudness $R\equiv L_{\nu}(6\ \mathrm{cm})/L_{\nu}(2500\ \mathrm{\AA})$ being 
greater than $10$) sources are removed. We reject light curves with less than $40$ 
epochs.\footnote{Only a small number of SDSS S82 
light curves have less than $40$ observations \citep[see Figure 2 of][]{Macleod2010}.} The 
resulting parent sample consists of $6271$ SDSS S82 quasars and will be used for subsequent 
variability modeling. The distributions of redshift, $M_{\mathrm{BH}}$, and $L_{\mathrm{3000}}$ 
for the parent sample are shown in Figure~\ref{fig:pms}. 

\begin{figure*}
    \plotone{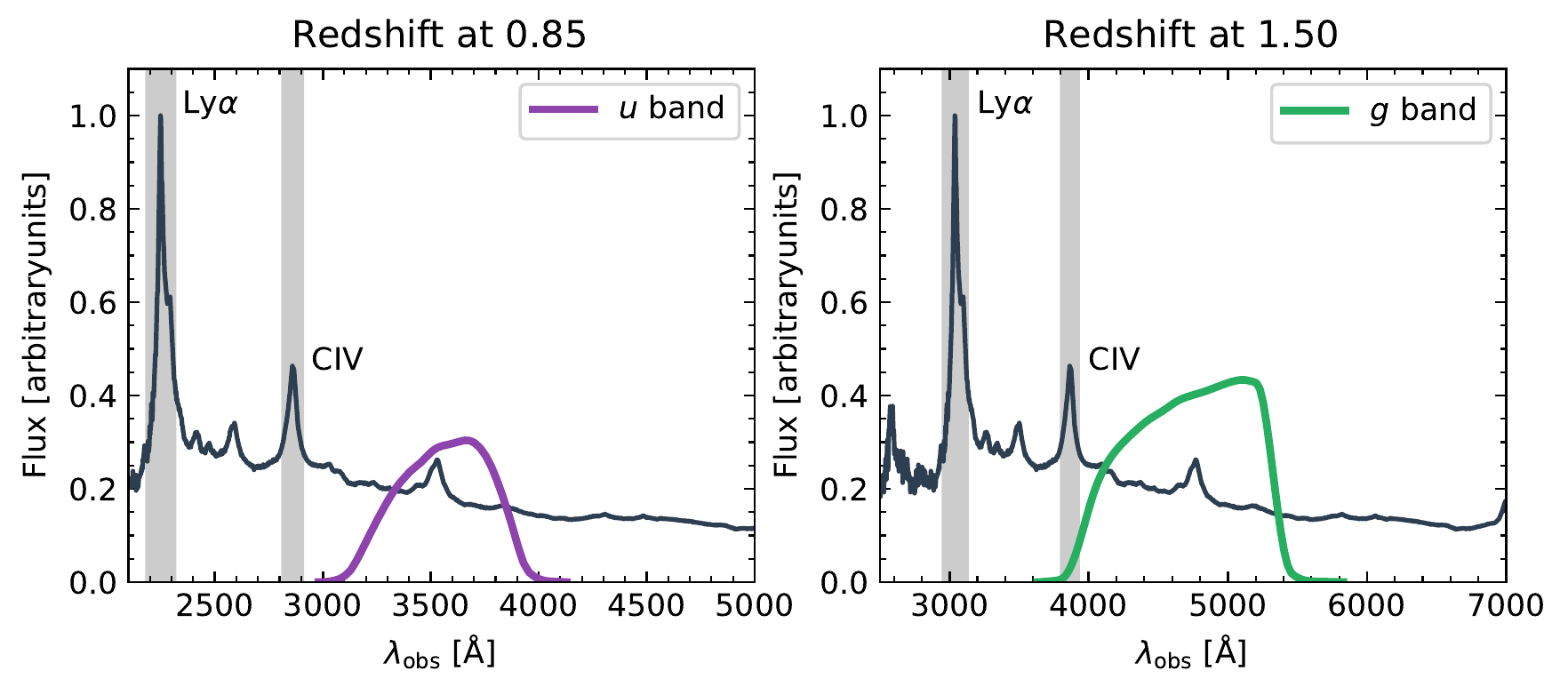}
    \caption{Illustration of redshift bins for defining the rest-frame continuum wavelengths. 
    Left: the \cite{Berk2001} composite SDSS spectrum is shown with the SDSS $u$-band filter 
    response curve for a quasar at $z=0.85$. Right: the same as the left panel but for a quasar 
    at $z=1.50$ and the SDSS $g$-band. If one considers higher-redshift $u$- or $g$-band light 
    curves, the contamination of strong broad emission lines (e.g., \CIV , Ly$\alpha$; 
    i.e., the shaded regions) is significant. \label{fig:filter}}
    \end{figure*}

\begin{figure*}
    \plotone{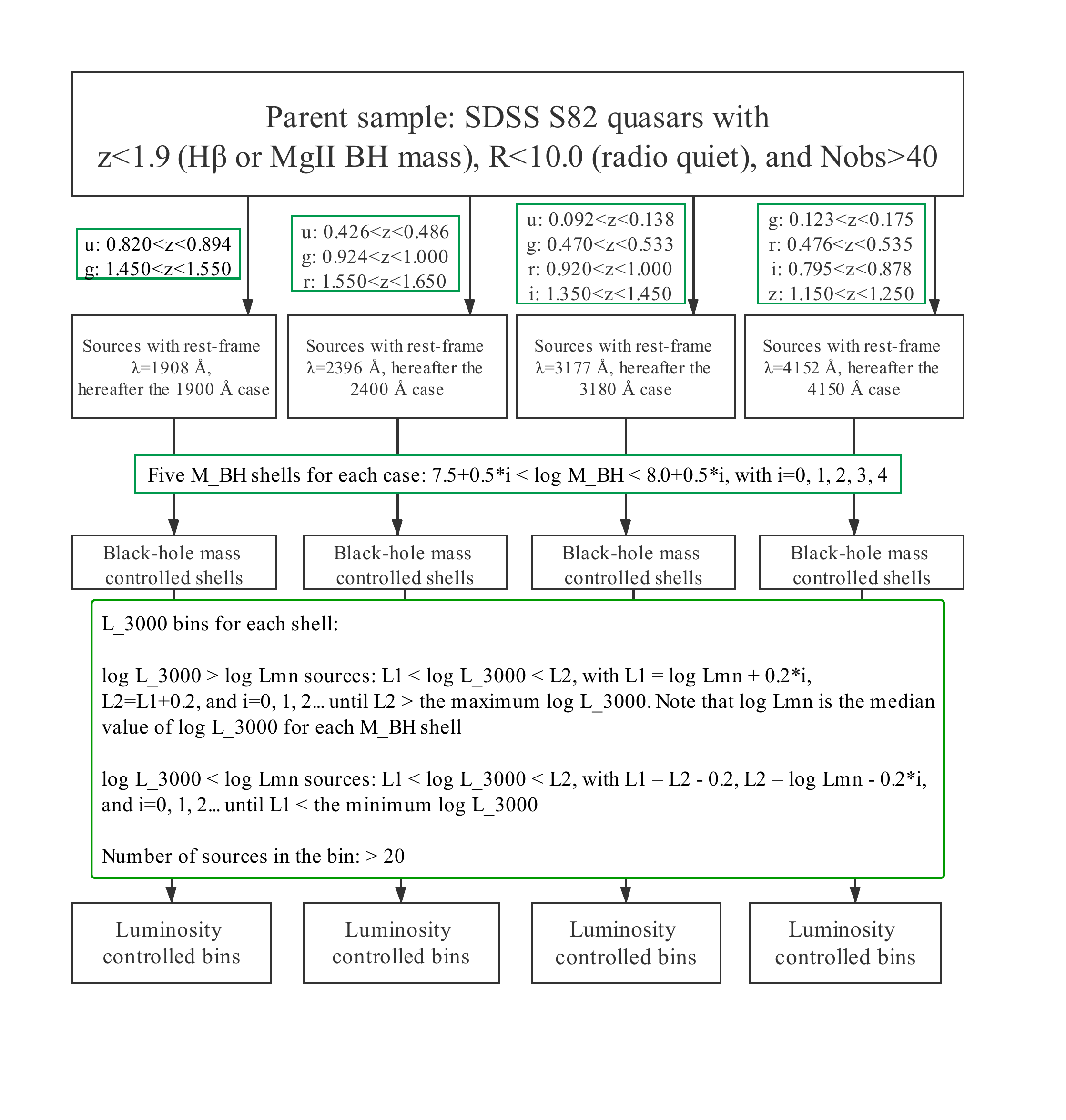}
    \caption{Our sample construction procedures to create $\lambda_{\mathrm{rest}}$, $M_{\mathrm{BH}}$, 
    and $L_{\mathrm{bol}}$ controlled bins. \label{fig:sample}}
    \end{figure*}

Previous studies \citep{Kelly2009,Macleod2010,Sun2015,Caplar2017,Sun2018a} show that quasar 
UV/optical variability depends at least on $M_{\mathrm{BH}}$, $L$, 
and rest-frame wavelength ($\lambda_{\mathrm{rest}}$). The wavelengths at which the $ugriz$ 
filters are the most sensitive are $3543\ \mathrm{\AA}$ ($u$ band), $4770\ \mathrm{\AA}$ ($g$ 
band), $6231\ \mathrm{\AA}$ ($r$ band), $7625\ \mathrm{\AA}$ ($i$ band), and $9134\ 
\mathrm{\AA}$ ($z$ band), respectively.\footnote{Please refer to 
\url{http://skyserver.sdss.org/dr1/en/proj/advanced/color/sdssfilters.asp}.} 
For a given band, the corresponding rest-frame wavelengths of observed AGNs depend on redshifts. 
We construct four rest-frame wavelength cases by 
following the redshift criteria in Figure~\ref{fig:sample}; for each case, these narrow redshift 
ranges are chosen to ensure that the corresponding rest-frame wavelengths for different SDSS 
filters are similar and to avoid strong broad emission-line coverage \citep[note that the 
broad emission lines are likely to be less variable than their nearby continua; see, 
e.g.,][]{Macleod2012, Sun2015}. Let us take the $1900\ \mathrm{\AA}$ case (i.e., the shortest 
$\lambda_{\mathrm{rest}} \simeq 4770/(1+1.5)=1908\ \mathrm{\AA}$ case) as an example. 
To obtain the shortest 
$\lambda_{\mathrm{rest}}$ light curves, we consider the $u$- and $g$-band observations 
of high-$z$ quasars. The $u$-band is sensitive to photons with $3000\ \mathrm{\AA}< 
\lambda_{\mathrm{obs}} < 4000\ \mathrm{\AA}$. Therefore, we should avoid using $u$-band light 
curves for redshifts higher than $0.9$; otherwise, the contamination of the prominent 
emission line, \CIV\ or Ly$\alpha$ , cannot be eliminated (see the left panel of 
Figure~\ref{fig:filter}). To ensure that $g$-band light curves probe the same 
$\lambda_{\mathrm{rest}}$ as the $u$-band ones, we should also avoid using the $g$-band light 
curves of $z>1.56$ sources (see the right panel of Figure~\ref{fig:filter}). We select the 
$u$-band light curves of $0.820<z <0.894$ quasars and the $g$-band light curves of $1.450<z<1.550$ 
quasars to probe the shortest $\lambda_{\mathrm{rest}} =1908\ \mathrm{\AA}$ (i.e., the $1900\ 
\mathrm{\AA}$ case) continuum variability. At these redshift ranges, \CIII\ is covered by $u$ 
and $g$ bands; however, \CIII\ is relatively weak with an equivalent width of $\sim 20\ 
\mathrm{\AA}$ \citep{Berk2001} and should have small contribution to the broad band fluxes. 
Other $\lambda_{\mathrm{rest}}$-controlled subsamples (i.e., the $2400\ \mathrm{\AA}$, 
$3180\ \mathrm{\AA}$, and $4150\ \mathrm{\AA}$ cases) are constructed following the same 
methodology (see Figure~\ref{fig:sample}). 

Each of the four $\lambda_{\mathrm{rest}}$-matched subsamples (i.e., the $1900\ \mathrm{\AA}$, 
$2400\ \mathrm{\AA}$, $3180\ \mathrm{\AA}$, and $4150\ \mathrm{\AA}$ cases) is divided 
into five ``shells'' according to $M_{\mathrm{BH}}$; and each shell has a $\log M_{\mathrm{BH}}$ 
width of $0.5$ dex \citep[i.e., the typical $1\sigma$ uncertainty of the virial $M_{\mathrm{BH}}$ 
estimators; for a review, see][]{Shen2013}, starting from $\log M_{\mathrm{BH}}=7.5$ (see the 
lower-left panel of Figure~\ref{fig:pms}). Each $M_{\mathrm{BH}}$ shell is further split into 
several $L_{\mathrm{3000}}$ bins following the methodology in Figure~\ref{fig:sample}. Note that 
only $L_{\mathrm{3000}}$ bins with more than $20$ sources will be considered in subsequent analyses. 
For each bin, we calculate the corresponding median $\log M_{\mathrm{BH}}$ and $\log L_{\mathrm{3000}}$. 
We then assume that all quasars in a bin have the same black-hole mass and bolometric luminosity 
(i.e., the median values) since the bolometric corrections (see Section~\ref{sect:mock}) and virial 
$M_{\mathrm{BH}}$ estimators are only valid in a sample-averaged sense. 

For each bin, we use the quasar light curves to calculate the ensemble structure function following 
the methodology of \cite{Macleod2012} and \cite{Sun2015}. That is, the statistical dispersion of two 
magnitude measurements ($\Delta m$) as a function of the corresponding rest-frame\footnote{Throughout 
this work, the wavelengths and timescales of quasar features are always rest-frame, unless otherwise 
specified.} time interval ($\Delta t$) is calculated as 
\begin{equation}
    \label{eq:sf}
    \mathrm{SF} = 0.74 \mathrm{IQR}(\Delta m) \\ ,
\end{equation}
where $\mathrm{IQR}(\Delta m)$ is the $25\%$--$75\%$ interquartile range of $\Delta m$; the constant 
$0.74$ normalizes the IQR to be equivalent to the standard deviation of a Gaussian distribution. 

For some sources, the light curves have a few outlier data points, i.e., a sudden increment of more 
than $1$ magnitude. To properly reject such outliers, we perform the following analysis. First, we 
use the CAR(1) model to fit each light curve. The CAR(1) model has been widely used to fit quasar 
UV/optical light curves\footnote{Note that data points with measurement errors $> 0.1$ mag are 
ignored.} \citep[e.g.,][]{Kelly2009,Macleod2010} although its validity on very short or long 
timescales has been questioned by several previous works \citep[e.g.,][]{Mushotzky2011, 
Caplar2017, Guo2017, Smith2018}. The CAR(1) model has the following covariance matrix
\begin{equation}
\label{eq:drw}
C_{i,j} = \frac{1}{2}\hat{\sigma}^2 \tau \exp{(-\Delta t_{i,j}/\tau)} \\,
\end{equation}
where $\hat{\sigma}$ and $\tau$ are the variability amplitude (i.e., standard deviation) of two 
observations ($i$ and $j$) on the timescale $\Delta t_{i,j}\equiv \left| t_i - t_j \right|=1$ day 
and the damping timescale in units of days, respectively. Second, we use a modified version of 
the \textit{Python} module \textit{qso\_fit.py}\footnote{This module can be accessed from 
http://astro.berkeley.edu/$\sim$nat/qso\_selection.} of \cite{Butler2011} to calculate the 
likelihood of a light curve given the CAR(1) model \citep[see Eq.~2 in][]{Butler2011}, and 
we find the combination of $\hat{\sigma}$ and $\tau$ that maximizes the likelihood.\footnote{We 
use the \textit{scipy} global optimization function, \textit{differential\_evolution}, to find 
the maximum likelihood.} Third, we use the best-fitting $\hat{\sigma}$ and $\tau$ to obtain 
the expected mean light curve and its statistical dispersion following \cite{Butler2011}. Fourth, 
we calculate the ratios of 
the absolute deviations between the observed and expected model light curves to the statistical 
dispersion for every epoch (hereafter the differential ratios). Fifth, we find the maximum value 
of the differential ratios; if the maximum differential ratio is larger than $3$, the corresponding 
epoch is rejected. The resulting light curve is then refitted with the CAR(1) model. We iterate this 
procedure until the maximum differential ratio is smaller than $3$. We stress that this process 
should not remove real (possibly non-CAR(1)) variability but only reject a few suspicious magnitude 
measurements. This is because the SDSS S82 quasar light curves are too sparse to distinguish 
between the CAR(1) model and other more complex stochastic processes \citep{Sun2018a}. 
These rejected data points often show unexpected strong flux variability ($\sim 0.3$ 
mag) within a few days or sharp flux changes ($\sim 1$ mag) on timescales of months. For 
more than $60\%$ of sources, no data point is rejected; for $\sim 25\%$ of sources, 
only one data point is rejected; for $\sim 10\%$ of sources, two data points are rejected; only 
for $\lesssim 5\%$ of sources, more than two data points (but less than six) are rejected. 
Therefore, this procedure should have negligible effect on our results. 

We can now use the light curves to obtain the corresponding ensemble structure 
functions. For a heterogeneous data sample, e.g., the quasar light-curve data from the Palomar 
Transient Factory, \cite{Caplar2017} point out that the observational data pairs at some specific 
$\Delta t$ might be dominated by a minor fraction of high-cadence sources; therefore, the resulting 
ensemble structure functions can be substantially biased. However, this bias should be not important 
in our case since 
the light-curve samplings of SDSS S82 are quite uniform. Therefore, we simply use all data pairs 
for a given $\Delta t$ to calculate the corresponding ensemble structure function. Note that the 
same strategy is adopted when calculating the model ensemble structure functions 
(see Section~\ref{sect:mock}) for the sake of consistency. 
The resulting ensemble structure functions for the $1900\ \mathrm{\AA}$, $2400\ \mathrm{\AA}$, 
$3180\ \mathrm{\AA}$, and $4150\ \mathrm{\AA}$ cases are shown in Figures~\ref{fig:sf-A}, 
\ref{fig:sf-B}, \ref{fig:sf-C}, and \ref{fig:sf-D}, respectively. We do not plot the uncertainties 
of the ensemble structure functions; the uncertainties will be assigned to the model results 
because we will consider the real sampling and measurement errors when generating mock light 
curves (see Section~\ref{sect:mock}). Note that the model-based error bars are consistent 
with the ones obtained from bootstrapping (with replacement) of quasars 
(i.e., similar to the procedure in \citealt{Caplar2017}).

\section{Modeling SDSS S82 quasar Multi-band Variability}
\label{sect:mock}
\subsection{Model I: A Constant Bolometric Correction}
\label{sect:start}
We use our CHAR model to simulate quasar UV/optical light curves. This model 
assumes that the corona and the underlying cold thin disk are magnetically coupled; coronal 
magnetic fluctuations can induce variations of the disk heating rate. The resulting disk 
effective temperature is calculated by considering the vertically integrated thermal-energy 
conservation law \citep[for more details, refer to Section 2 of][]{Sun2020}. The free parameters 
of the CHAR model are black-hole mass, absolute accretion rate ($\dot{M}$), the dimensionless 
viscosity parameter $\alpha$, and the variability amplitude of the heating rate 
($\delta_{\mathrm{mc}}$). 

For each bin, we use $\dot{M}=L_{\mathrm{bol}}/(\eta c^2)$ with the radiative efficiency $\eta= 
0.1$ to estimate $\dot{M}$,\footnote{We do not use $L_{3000}$ and the thin-disk 
theory to infer $\dot{M}$ because the inclination angle and the intrinsic extinction are unknown. 
Therefore, we prefer to adopt the empirical bolometric corrections to estimate 
$\dot{M}$.} where $L_{\mathrm{bol}}=5.15 L_{\mathrm{3000}}$ \citep{Richards2006}; 
then, the model $M_{\mathrm{BH}}$ and $\dot{M}$ are both fixed to be the observed ones. The 
remaining free parameters are $\alpha$ and $\delta_{\mathrm{mc}}$. We use the lowest 
$L_{\mathrm{bol}}$ bin of the $2400\ \mathrm{\AA}$ case \footnote{We choose this case because the 
light curves are mostly from $r$-band observations whose measurement uncertainties are the smallest 
among the five filters.} to determine $\alpha$ and $\delta_{\mathrm{mc}}$. That is, we consider 
the combinations of $\alpha$ and $\delta_{\mathrm{mc}}$ by stepping through 14 values of 
$\delta_{\mathrm{mc}}$ from $0.1$ to $0.7$ in equal linear increments and $20$ values of $\alpha$ 
from $10^{-2}$ to $10^{-0.2}$ in equal logarithmic increments. For each source in the bin, we 
use the same $M_{\mathrm{BH}}$, $\dot{M}$, $\delta_{\mathrm{mc}}$, and $\alpha$ as the CHAR model 
parameters to simulate the same number of mock light curves; the mock light curves are shifted to 
the observed frame according to their redshifts; the sampling patterns of the mock 
light curves are the same as the observed ones. We then add measurement noise to each mock light 
curve using uncorrelated white noise whose variance is the same as the observed one 
which is estimated from the observed ensemble structure functions at small $\Delta 
t$ (i.e., $\Delta t< 4$ days). Subsequently, 
we calculate the mock ensemble structure function by using the mock light curves. We repeat the 
simulation $50$ times. The differences between the mean results of the $50$ mock ensemble structure 
functions and the observed ones for all combinations of $\alpha$ and $\delta_{\mathrm{mc}}$ are 
calculated. The best-fitting combination of $\alpha$ and $\delta_{\mathrm{mc}}$ (hereafter, $\alpha 
(\mathrm{best})$ and $\delta_{\mathrm{mc}}(\mathrm{best})$) is the one that minimizes the differences 
between the observed and the model ensemble structure functions. We find that $\alpha (\mathrm{best}) 
= 0.5$ and $\delta_{\mathrm{mc}}(\mathrm{best})=0.5$ (i.e., the structure function of the natural 
logarithmic heating rate fluctuation on the timescale of $100$ days is $0.5$). 

To model the ensemble structure functions of the rest of the bins, we fix $\alpha$ and 
$\delta_{\mathrm{mc}}$ to be $\alpha (\mathrm{best})$ and $\delta_{\mathrm{mc}}(\mathrm{best})$ 
determined above, respectively. We only change $\log M_{\mathrm{BH}}$ and $\log L_{\mathrm{bol}}$ 
according to the observed values, i.e., \textit{there is no free parameter in the following modeling 
procedures}. For the rest of the bins of each case, we generate the corresponding mock light curves 
by following the same procedure mentioned above. Again, the real sampling patterns and measurement 
noise are taken into consideration. We repeat this process $400$ times (i.e., for each source, 
$400$ mock light curves with the same cadence and measurement noise are generated). 

The mock ensemble structure functions for the $1900\ \mathrm{\AA}$, $2400\ \mathrm{\AA}$, $3180\ 
\mathrm{\AA}$ and $4150\ \mathrm{\AA}$ cases are shown in Figures~\ref{fig:sf-A}, 
\ref{fig:sf-B}, \ref{fig:sf-C}, and \ref{fig:sf-D}, respectively. Just like the observed ensemble 
structure functions, our mock ensemble structure functions depend weakly on $M_{\mathrm{BH}}$ (see 
each of the row panels in Figures~\ref{fig:sf-A}, \ref{fig:sf-B}, \ref{fig:sf-C}, and \ref{fig:sf-D}) 
but highly anti-correlate with $L_{3000}$ and $\lambda_{\mathrm{rest}}$ (see Figure~\ref{fig:sf-pms} 
and the column panels in Figures~\ref{fig:sf-A}, \ref{fig:sf-B}, \ref{fig:sf-C}, and \ref{fig:sf-D}). 
That is, without fine-tuning of the model parameters, our mock ensemble structure functions are 
broadly consistent with the observed ones. 

To quantitatively assess our modeling results, we calculate the following statistic for each bin, 
\begin{equation}
    \label{eq:rchi2}
    \hat{S} = \mathrm{Median} \Bigg(\frac{(\overline{\log_{10} \mathrm{SF}(\Delta t)}-\log_{10} 
    \mathrm{SF}_{\mathrm{obs}}(\Delta t))^2}{\sigma_{\mathrm{SF}}^2} \Bigg) \\,
\end{equation}
where $\overline{\log_{10} \mathrm{SF}(\Delta t)}$, $\sigma_{\mathrm{SF}}$, and $\log_{10} 
\mathrm{SF}_{\mathrm{obs}}(\Delta t)$ are the mean of the decimal logarithm of the $400$ model ensemble 
structure functions, its $1\sigma$ uncertainty (i.e., the standard deviation of the $400$ model ensemble 
structure functions), and the decimal logarithm of the observed ensemble structure function, respectively. 
Note that only $\Delta t>10\ \mathrm{days}$ are considered since measurement noise dominates over quasar 
variability on shorter time intervals.\footnote{For the $4150\ \mathrm{\AA}$ case, only data points with 
$\Delta t>20\ \mathrm{days}$ are considered. We use a larger $\Delta t$ cut because the intrinsic 
variability of this case is the smallest among the four cases.} Our definition of $\hat{S}$ is similar 
to the traditional reduced $\chi^2$ statistic, but is more robust against outliers. The modeling statistic 
for each case is then defined as 
\begin{equation}
    \label{eq:rtot}
    \hat{S}_{\mathrm{tot}} = \sum_{i=1}^{N} \hat{S}_i \\,
\end{equation}
where $N$ is the number of bins in each case. The expected distribution of the statistic 
$\hat{S}_{\mathrm{tot}}$ is unknown because the adjacent SF estimates are correlated and the light curves 
have irregular gaps.\footnote{This argument is also valid if we adopt the traditional $\chi^2$ statistic. 
For a detailed discussion of this point, refer to \cite{Emmanoulopoulos2010} and references therein.} 
However, we can infer the distribution by using simulations. That is, we use Eqs.~\ref{eq:rchi2} and 
\ref{eq:rtot} to obtain the mock $\hat{S}_{\mathrm{tot}}$ (hereafter $\hat{S}_{\mathrm{tot,mc}}$) for 
each of the $400$ mock ensemble structure functions (i.e., replacing $\mathrm{SF}_{\mathrm{obs}}$ in 
Eq.~\ref{eq:rchi2} with a mock ensemble structure function). Then, for each case, we find that the 
histogram of $\hat{S}_{\mathrm{tot,mc}}$ can be described by a Gamma 
distribution. Hence, we fit each distribution of $\hat{S}_{\mathrm{tot,mc}}$ with a Gamma distribution. 
We use the Kolmogorov–Smirnov test to justify our best-fitting distribution and confirm that the null 
hypothesis (i.e., the best-fitting Gamma distribution is consistent with the observed one) cannot be 
rejected. 

For each case, we use the best-fitting Gamma distribution to calculate the following statistical 
parameters, i.e., the probability of $\hat{S}_{\mathrm{tot,mc}}>\hat{S}_{\mathrm{tot}}$ (hereafter 
$p_0$), the natural logarithm likelihood of $\hat{S}_{\mathrm{tot,mc}}=\hat{S}_{\mathrm{tot}}$ 
(hereafter $\ln L_0$), and the Akaike information criterion \citep[AIC;][]{Akaike1974}. The AIC is 
defined as follows: 
\begin{equation}
\label{eq:aic}
\mathrm{AIC} = 2f - 2\ln L_0 \\,
\end{equation}
where $f=4$ is the number of model parameters. The values of the three statistical parameters are 
listed in Table~\ref{tab:sts}. 

For the $1900\ \mathrm{\AA}$ and $2400\ \mathrm{\AA}$ cases the corresponding $p_0$ values are much 
larger than $0.01$; that is, at a significance level of $0.01$, we cannot reject the null 
hypothesis that the differences between our mock ensemble structure functions and the observed ones 
are due to statistical fluctuations. For the $3180\ \mathrm{\AA}$ and $4150\ \mathrm{\AA}$ cases, 
their $p_0$ values are close than to even smaller than $0.01$; we argue that this deviation is 
because the galaxy stellar light dilutes the $\lambda_{\mathrm{rest}}=3180\ \mathrm{\AA}$ and 
$\lambda_{\mathrm{rest}}=4150\ \mathrm{\AA}$ emission variability. Indeed, the differences between 
the mock and observed ensemble structure functions are prominent only in those $L_{\mathrm{bol}}< 
10^{46}\ \mathrm{erg\ s^{-1}}$ bins (see Figure~\ref{fig:sf-D}). All in all, we conclude that the 
CHAR model can satisfactorily reproduce the dependence of quasar UV/optical variability upon 
$M_{\mathrm{BH}}$, $L_{\mathrm{bol}}$, and $\lambda_{\mathrm{rest}}$, without any fine-tuning. 

\begin{deluxetable*}{ccccccc}
    \tablecaption{The statistical parameters for the two models. \label{tab:sts}}
    \tablehead{\colhead{Statistical parameters$^a$} & \colhead{the $1900\ \mathrm{\AA}$ case} & 
    \colhead{the $2400\ \mathrm{\AA}$ case} & \colhead{the $3180\ \mathrm{\AA}$ case} & 
    \colhead{the $4150\ \mathrm{\AA}$ case}}
    \startdata
    $p_0$ & $5.62\times 10^{-2}$   & $6.77\times 10^{-2}$ & $1.94\times 10^{-3}$  & $1.63\times 10^{-2}$  \\ 
    $p_1$ &  $2.83\times 10^{-7}$  & $1.03\times 10^{-4}$  & $2.70\times 10^{-2}$   & $8.84\times 10^{-3}$  \\ 
    \hline
    $\ln L_0$ & $-0.56$   & $-0.37$  & $-3.52$  & $-1.60$  \\ 
    $\ln L_1$ & $-12.23$   & $-6.29$  & $-1.32$  & $-2.23$  \\ \hline
    $\mathrm{AIC}_0$ & $9.12$   & $8.74$   & $15.04$  & $11.21$  \\ 
    $\mathrm{AIC}_1$ & $32.46$  & $20.59$  & $10.65$   & $12.47$ \\
    \enddata
    \tablenotetext{a}{Subscripts $0$ and $1$ refer to models I and II, respectively.}
\end{deluxetable*}

\subsection{Model II: A Luminosity-dependent Bolometric Correction}
\label{sect:alt}
Our model light curves are sensitive to $M_{\mathrm{BH}}$ and $L_{\mathrm{bol}}$ (or $\dot{M}$). 
There is growing evidence that the current virial black-hole mass estimators (using either 
H$\beta$ or \MgII) suffer from significant systematic biases \citep[e.g.,][]{Grier2017, Du2018, 
Fonseca2020}. The accuracy 
of $M_{\mathrm{BH}}$ estimation might be greatly improved in ongoing or future RM campaigns 
\citep[e.g., SDSS-RM; see][]{Shen2016, Grier2017}. Meanwhile, some previous works also 
suggested that the bolometric corrections of the rest-frame $3000\ \mathrm{\AA}$ or $5100\ 
\mathrm{\AA}$ are not constant but depend on $L_{\mathrm{bol}}$ \citep[e.g., ][]{Nemmen2010, 
Netzer2019}. For instance, \cite{Nemmen2010} calculated the spectral energy distributions (SEDs) 
of thin accretion disks with various $M_{\mathrm{BH}}$, $L_{\mathrm{bol}}$, and inclination 
angles and found the following alternative bolometric correction: 
\begin{equation}
\label{eq:bolcor}
\log L_{\mathrm{bol}} = C_1 + C_2 \log L_{3000} \\,
\end{equation}
where $C_1 = 9.24$ and $C_2=0.81$. An almost identical relation was obtained by \cite{Netzer2019}. 
We also try to use this bolometric correction to estimate $L_{\mathrm{bol}}$ (or $\dot{M}$) and 
repeat the modeling procedures in Section~\ref{sect:start} to obtain the new mock ensemble structure 
functions. 

\begin{figure}
    \plotone{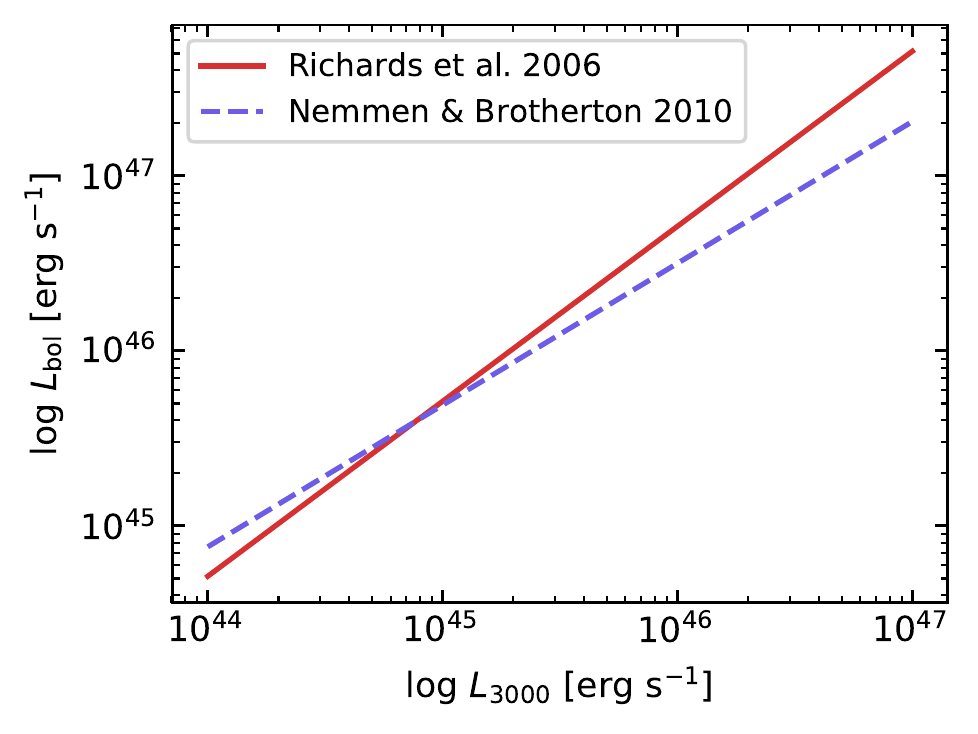}
    \caption{The constant and the luminosity-dependent bolometric corrections. For luminous 
    sources, the bolometric correction of \cite{Nemmen2010} is smaller than the constant one. 
    \label{fig:bol}}
\end{figure}

Compared with model I, the mock ensemble structure functions of model II for high-luminosity 
bins have larger variability amplitudes. This is simply because, for high-luminosity bins, 
the bolometric correction of \cite{Nemmen2010} is smaller than (see Figure~\ref{fig:bol}) 
that of \cite{Richards2006}. 

We can also calculate the following three statistical parameters, i.e., the probability of 
$\hat{S}_{\mathrm{tot,mc}}>\hat{S}_{\mathrm{tot}}$ (hereafter $p_1$), the natural logarithm 
likelihood of $\hat{S}_{\mathrm{tot,mc}}=\hat{S}_{\mathrm{tot}}$ (hereafter $\ln L_1$), and 
the AIC (hereafter $\mathrm{AIC}_1$) for model II. Their values are also listed 
in Table~\ref{tab:sts}. For all but one cases, we can reject the null hypothesis that the 
mock ensemble structure functions are consistent with the observed ones at a significance 
level of $0.01$; for the $3180\ \mathrm{\AA}$ case, its $p_0$ value is sightly larger than 
$0.01$. The total AIC of model II is also larger than that of model I with a difference 
of $32.06$. Hence, we can conclude that model I with the constant bolometric correction of 
\cite{Richards2006} is favored over model II with the luminosity-dependent bolometric 
correction of \cite{Nemmen2010}. Our result is consistent with some independent works 
\citep[e.g.,][]{Runnoe2012, Duras2020} who also found that the bolometric correction for 
$L_{3000}$ is constant over seven luminosity decades.

\section{Discussion}
\label{sect:dis}
As demonstrated in previous sections and \cite{Sun2020}, our CHAR model has the potential 
to satisfactorily explain many aspects of quasar UV/optical variability, including its 
dependence upon quasar physical properties. This is because the ratio of the observed to 
thermal timescales ($\tau_{\mathrm{TH}}$) almost determines the variability behavior 
(i.e., the variability behavior is nearly $\tau_{\mathrm{TH}}$-scale-invariant; see Section 
2.2 of \citealt{Sun2020}), and the thermal timescale scales as $\alpha^{-1} \dot{M}^{0.5} 
\lambda_{\mathrm{rest}}^2$. If we increase $L_{\mathrm{bol}}$, the thermal timescale 
also increases since $\dot{M}=L_{\mathrm{bol}}/(\eta c^2)$ and the variability amplitude 
decreases. Hence, our CHAR model provides a natural explanation of the dependence of 
quasar UV/optical variability upon $L_{\mathrm{bol}}$, and our modeling results are able 
to distinguish between the constant and luminosity-dependent bolometric corrections. 

Just like the luminosity-dependent bolometric corrections of \cite{Nemmen2010} and 
\cite{Netzer2019}, our CHAR model is largely based on the classical thin disk model 
\citep[SSD;][]{SSD}. Thus, why do our CHAR modeling results favor a constant bolometric 
correction? We speculate that this is because, as pointed out by \cite{Netzer2019}, 
the unknown parameters, e.g., SMBH spin and sightline, can introduce significant 
uncertainties to the bolometric corrections; hence, the performance of the SSD-based 
bolometric corrections is worse than that of the simple constant correction. 

According to our CHAR model, for fixed frequency ranges (or fixed time samplings), the 
shapes of the power spectral densities (PSDs) of low-luminosity sources are flatter than 
their high-luminosity counterparts. The physical reasons are as follows. 
First, the shape of our model PSD is almost the same if the frequency is expressed in 
units of $\tau_{\mathrm{TH}}$ (see Section~2.2 and Figure~15 of \citealt{Sun2020}). 
On timescales much less than $\tau_{\mathrm{TH}}$, the disk temperature cannot respond 
to the fluctuations of the heating rate and the variations of the blackbody disk emission 
are suppressed, i.e., the PSD declines steeply at high frequencies (i.e., small 
$\Delta t/\tau_{\mathrm{TH}}$). On long timescales (comparable to or larger than 
$\tau_{\mathrm{TH}}$), the disk temperature can vary in response to the fluctuations of 
the heating rate and the variations of the disk emission are preserved. 
Second, the thermal timescale $\tau_{\mathrm{TH}}$ scales as $\dot{M}^{0.5}\lambda^2$. 
For fixed observational timescales ($\Delta t$), high-luminosity sources have small 
$\Delta t/\tau_{\mathrm{TH}}$ and their light curves can only probe the steep parts 
of the PSDs. On the other hand, low-luminosity sources have large $\Delta t/\tau_{\mathrm{TH}}$ 
and their light curves can probe the flat parts of the PSDs. Such a dependence is found 
for the SDSS S82 quasars by \cite{Caplar2017}. Very recently, \cite{Burke2020} used the 
optical light curve 
of NGC 4395 from the Transiting Exoplanet Survey Satellite to probe its hours-to-weeks optical 
variability and found that the PSD is consistent with that of the CAR (1) model. This result 
seems to be incompatible with the Kepler observations \citep{Mushotzky2011}. Our CHAR model 
provides a natural explanation for this apparent inconsistency. The bolometric luminosity 
of NGC 4395 is fainter than that of the best-studied \textit{Kepler} AGN Zw 229-15 by a factor 
of $10^3$, i.e., (for fixed 
$\alpha$ and $\lambda_{\mathrm{rest}}$) the thermal timescale of the former is $10^{1.5} 
\sim 30$ times shorter than that of the latter (since the thermal timescale $\propto 
\dot{M}^{0.5}$). As shown by \cite{Kelly2014}, the PSD of Zw 229-15 approaches 
the $f^{-2}$ relation on timescales $\geq 10$ days (and our CHAR model indeed reproduces 
this behavior; see Figure 18 of \citealt{Sun2020}). Therefore, according to our CHAR model, 
on timescales $\geq 10/30=1/3$ days (i.e., $f\leq 3\ \mathrm{day}^{-1}$), the PSD of NGC 4395 
is also expected to follow the $f^{-2}$ relation. 

For fixed $L_{\mathrm{bol}}$ and $\lambda_{\mathrm{rest}}$, our CHAR model predicts that 
quasar UV/optical variability amplitude increases slightly with $M_{\mathrm{BH}}$ (see 
Figure~13 of \citealt{Sun2020}); this prediction is also consistent with SDSS S82 
observations (see the row panels in Figures~\ref{fig:sf-A}, \ref{fig:sf-B}, \ref{fig:sf-C}, 
and \ref{fig:sf-D}). For fixed $M_{\mathrm{BH}}$ and $L_{\mathrm{bol}}$, the thermal 
timescale correlates with $\lambda_{\mathrm{rest}}$; therefore, the UV continuum is more 
variable than the optical one (see Figure~\ref{fig:sf-pms}). 

When modeling the ensemble structure functions, $\alpha$ is fixed for all sources. If 
$\alpha$ is allowed to decrease with increasing $L_{\mathrm{bol}}$, we can also reproduce 
the observed ensemble structure functions with the luminosity-dependent bolometric correction 
(see Eq.~\ref{eq:bolcor}). However, since the constant bolometric correction for $L_{3000}$ 
is also favored in other independent works \citep[e.g.,][]{Runnoe2012, Duras2020}, our results 
indicate that the assumption of a constant $\alpha$ (i.e., $\alpha$ should depend at most weakly 
upon $M_{\mathrm{BH}}$ and $L_{\mathrm{bol}}$) is probably reasonable. This conclusion is 
further supported by the similar $\alpha$ values found in the accretion disks around stellar 
black holes (whose $M_{\mathrm{BH}}$ and $L_{\mathrm{bol}}$ are several orders of magnitude 
smaller than AGNs) in their outburst phases \citep{King2007}. Note that, for fixed 
$L_{\mathrm{bol}}$, $\dot{M} \propto \eta^{-1}$ and $\eta$ correlates with dimensionless SMBH 
spin parameter ($a_*$, which takes values from $-1$ to $1$). In previous sections, we assume 
$\eta=0.1$, which corresponds to a moderate positive $a_*$. If the SDSS S82 SMBHs spin faster 
and have larger $\eta$, the inferred $\dot{M}$ is smaller and so is the required $\alpha$. 
Indeed, current X-ray spectroscopic observations \citep[for a recent review, see][]{Reynolds2019} 
seem to find a large fraction of SMBHs with $a_* > 0.9$ \citep[but see][]{Laor2019}. 

There are still some small residuals between the two models and the observed ensemble structure 
functions. We test the possible correlations between the small residuals and quasar physical 
properties (i.e., $M_{\mathrm{BH}}$ and $L_{\mathrm{3000}}$) and find that the correlations 
are statistically insignificant (i.e., the corresponding $p_0$ values are much greater than 
$0.01$). We speculate that the small residuals are driven by the significant uncertainties 
of $M_{\mathrm{BH}}$ \cite[for instance, while the $M_{\mathrm{BH}}$ estimators depend upon 
orientation, the variability amplitude should be insensitive to orientation; see, 
e.g.,][]{Sun2018a} and $\dot{M}$. It is also possible that quasar UV/optical variability might 
also depend (weakly) upon other additional factors, e.g., X-ray loudness \citep{Kang2018} or 
magnetic field \citep{Cai2019}.

\section{Summary and Future Work}
\subsection{Summary}
\label{sect:sum}
We use our CHAR model to reproduce SDSS S82 quasar UV/optical variability. Our main results are 
summarized as follows: 
\begin{itemize}
    \item The CHAR model can broadly reproduce the observed ensemble structure functions of SDSS S82 
    quasars with various $M_{\mathrm{BH}}$, $L_{\mathrm{bol}}$, and $\lambda_{\mathrm{rest}}$ without 
    fine-tuning the model parameters. 
    \item Our variability modeling results are in favor of a constant bolometric correction for the 
    $3000\ \mathrm{\AA}$ continuum luminosity. 
    \item The dimensionless viscosity parameter $\alpha$ should depend only weakly on $M_{\mathrm{BH}}$ 
    and $L_{\mathrm{bol}}$. 
    \item Based on our physical modeling results, we present a recipe to simulate AGN UV/optical light 
    curves. 
\end{itemize}
Compared to empirical-model fitting results, our results demonstrate a new way to directly infer 
quasar properties (e.g., the bolometric correction, the dimensionless viscosity dimensionless parameter) 
by physically modeling their multi-wavelength light curves. 

\subsection{Future Work}
\label{sect:future}
Future time-domain surveys like the Legacy Survey of Space and Time (LSST) will provide well-sampled 
(with a cadence of $\sim 3$ days) light curves for $\gtrsim 10^4$ AGNs \citep{Brandt2018, Scolnic2018}. 
Combining the LSST light curves with archival observations or the sparse but deep multi-band 
observations of the Chinese Space Station Telescope \citep{Zhan2011, Cao2018}, the light curves can 
cover an observed-frame timescale of over $30$ years. In the future, we will use the CHAR model 
to simulate mock LSST AGN light curves\footnote{The authors are willing to share the simulated 
light curves upon request.} and explore AGN UV/optical variability on very long timescales 
\citep[e.g.,][]{Macleod2012,Caplar2020}. 

\acknowledgments
We thank the referee, Dr.~Neven Caplar, for his helpful comments that improved the manuscript.
MYS acknowledges support from the National Natural Science Foundation of China (NSFC-11973002, 
NSFC-11873045). 
YQX and JXW acknowledge support from the National Natural Science Foundation of China (NSFC-11890693, 
11421303), the CAS Frontier Science Key Research Program (QYZDJ-SSW-SLH006), and the K.C. Wong Education 
Foundation. 
WNB acknowledges support from NSF grant AST-1516784 and NASA ADAP grant 80NSSC18K0878. 
TL acknowledges support from the National Natural Science Foundation of China (NSFC-11822304).

\software{Matplotlib \citep{Hunter2007}, Numpy \& Scipy \citep{scipy}, pyLCSIM \citep{Campana2017}}

\begin{figure*}
    \epsscale{0.9}
    \plotone{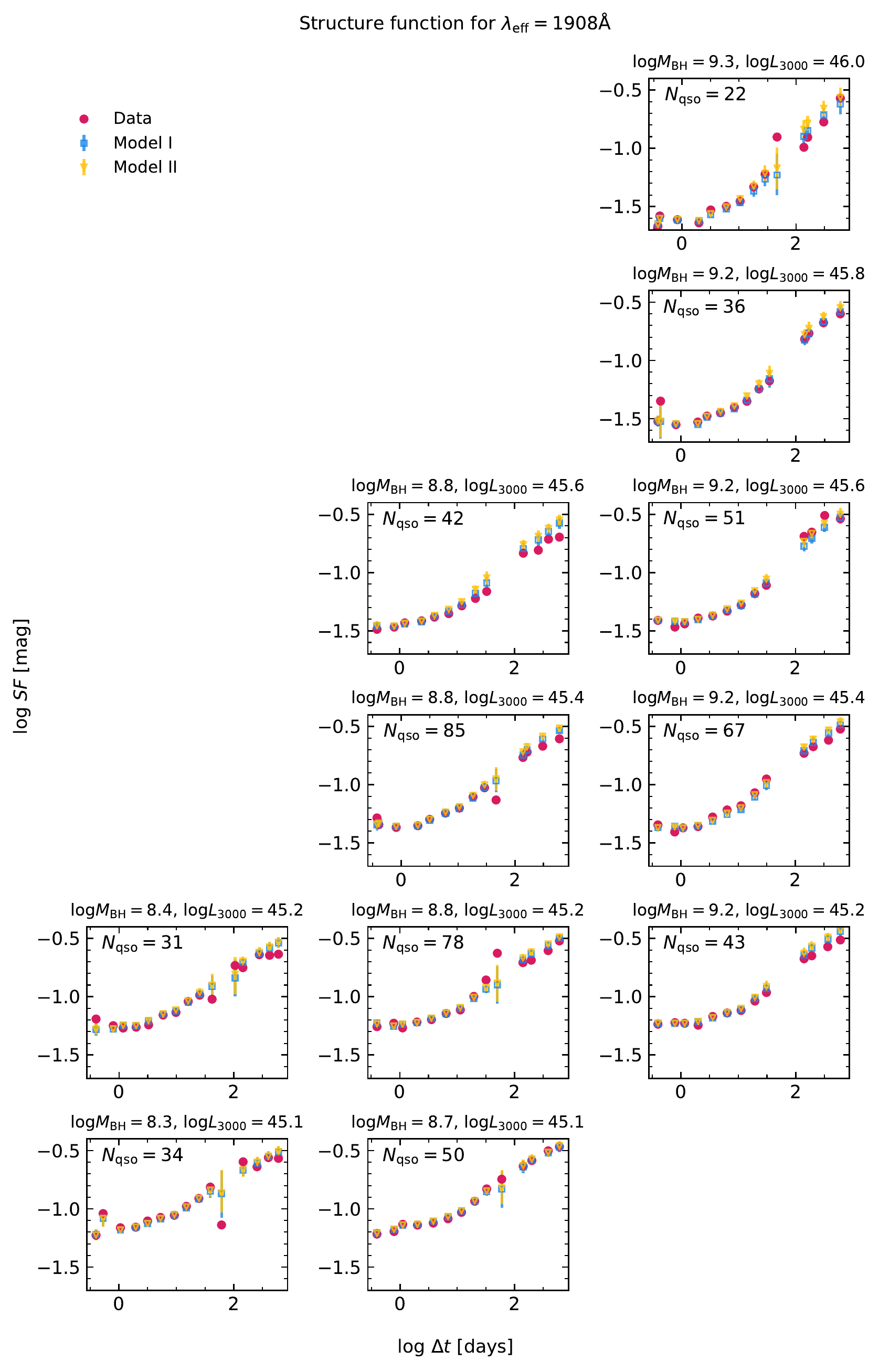}
    \caption{The ensemble structure functions for different [$M_{\mathrm{BH}}$, $L_{3000}$] 
    bins of the $1900\ \mathrm{\AA}$ case. Panels in each column (row) share the same $M_{\mathrm{BH}}$ 
    ($L_{3000}$). 
    The data is obtained from the SDSS S82 quasar light curves. The only difference in models I and 
    II is about the bolometric corrections used to convert $L_{3000}$ 
    into $L_{\mathrm{bol}}$. In model I, we use a constant bolometric correction of $5.15$; in model II, 
    a luminosity-dependent bolometric correction of \cite{Nemmen2010} is adopted. Note that a few 
    structure function data points show strong fluctuations which is simply caused by sampling issues. 
    The error bars indicate the standard deviations of the 400 model ensemble structure functions of Model 
    I and II. The time interval $\Delta t$ is in rest-frame. The gap around the rest-frame $100$ days 
    is caused by the lack of timescale coverage of SDSS S82 quasar light curves. 
    \label{fig:sf-A}}
\end{figure*}

\begin{figure*}
    \plotone{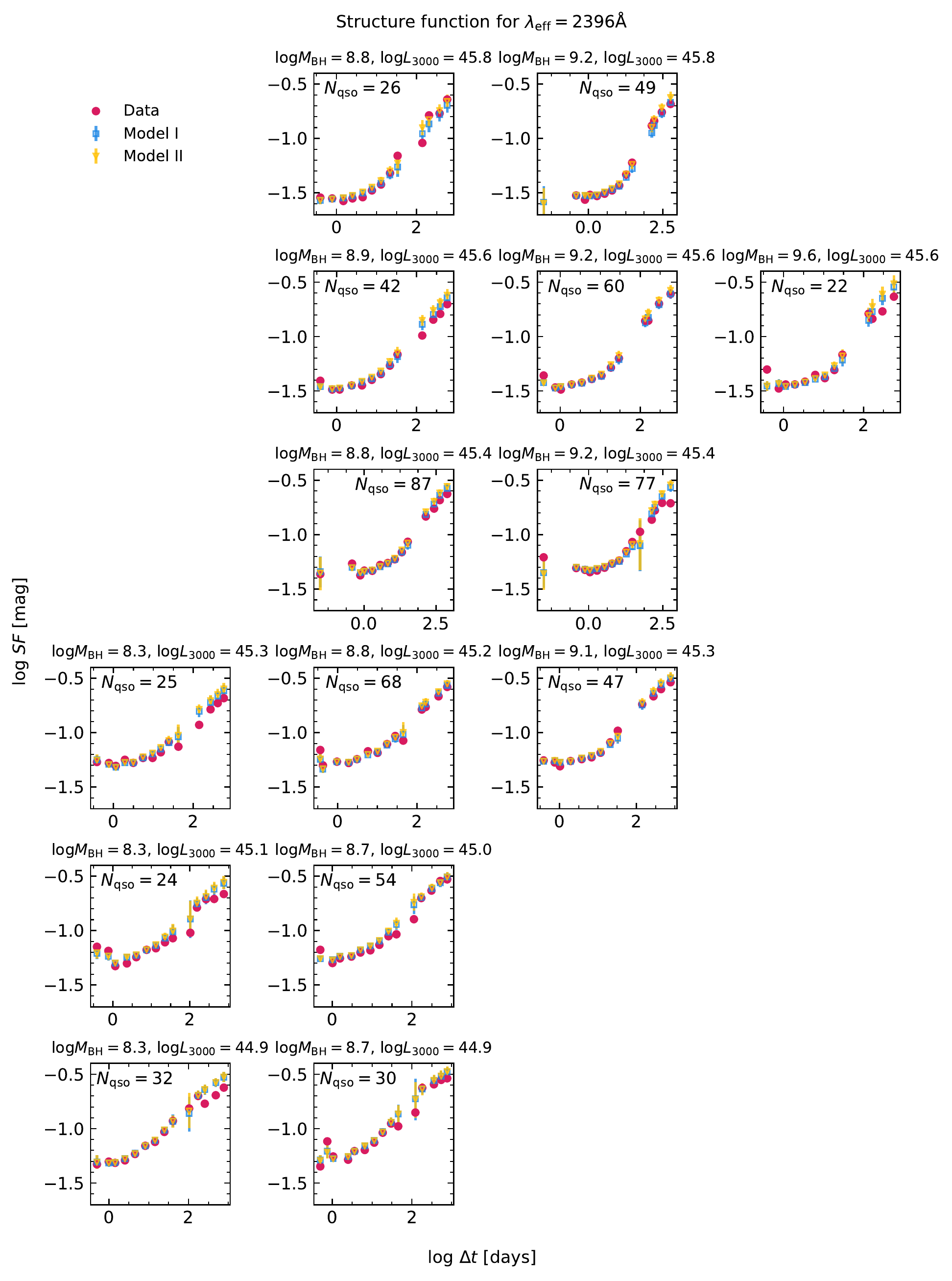}
    \caption{The ensemble structure functions for different [$M_{\mathrm{BH}}$, $L_{3000}$] 
    bins of the $2400\ \mathrm{\AA}$ case. Panels in each column (row) share the same $M_{\mathrm{BH}}$ 
    ($L_{3000}$). 
    \label{fig:sf-B}}
\end{figure*}

\begin{figure*}
    \epsscale{0.8}
    \plotone{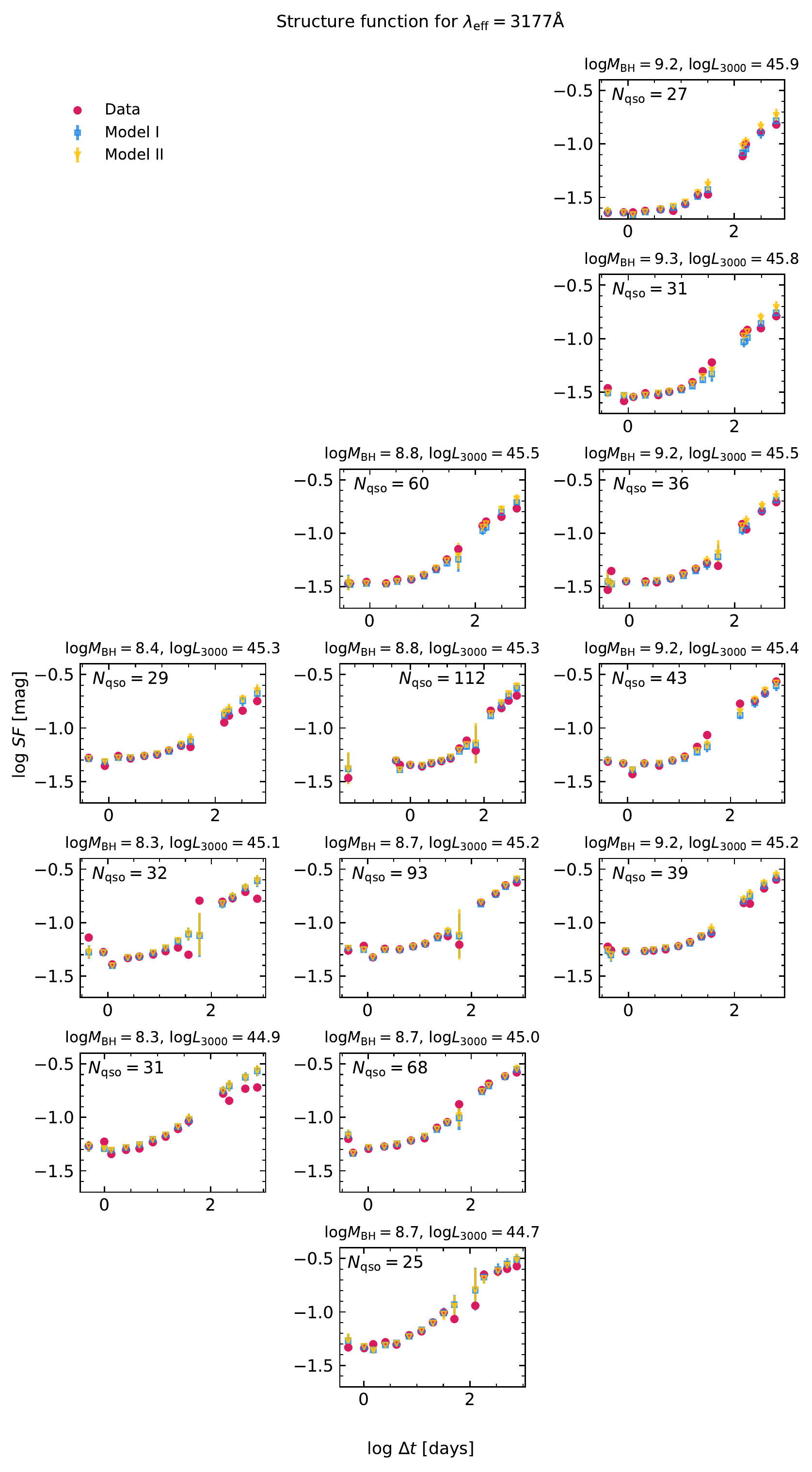}
    \caption{The ensemble structure functions for different [$M_{\mathrm{BH}}$, $L_{3000}$] 
    bins of the $3180\ \mathrm{\AA}$ case. Panels in each column (row) share the same $M_{\mathrm{BH}}$ 
    ($L_{3000}$). 
    \label{fig:sf-C}}
\end{figure*}

\begin{figure*}
    \plotone{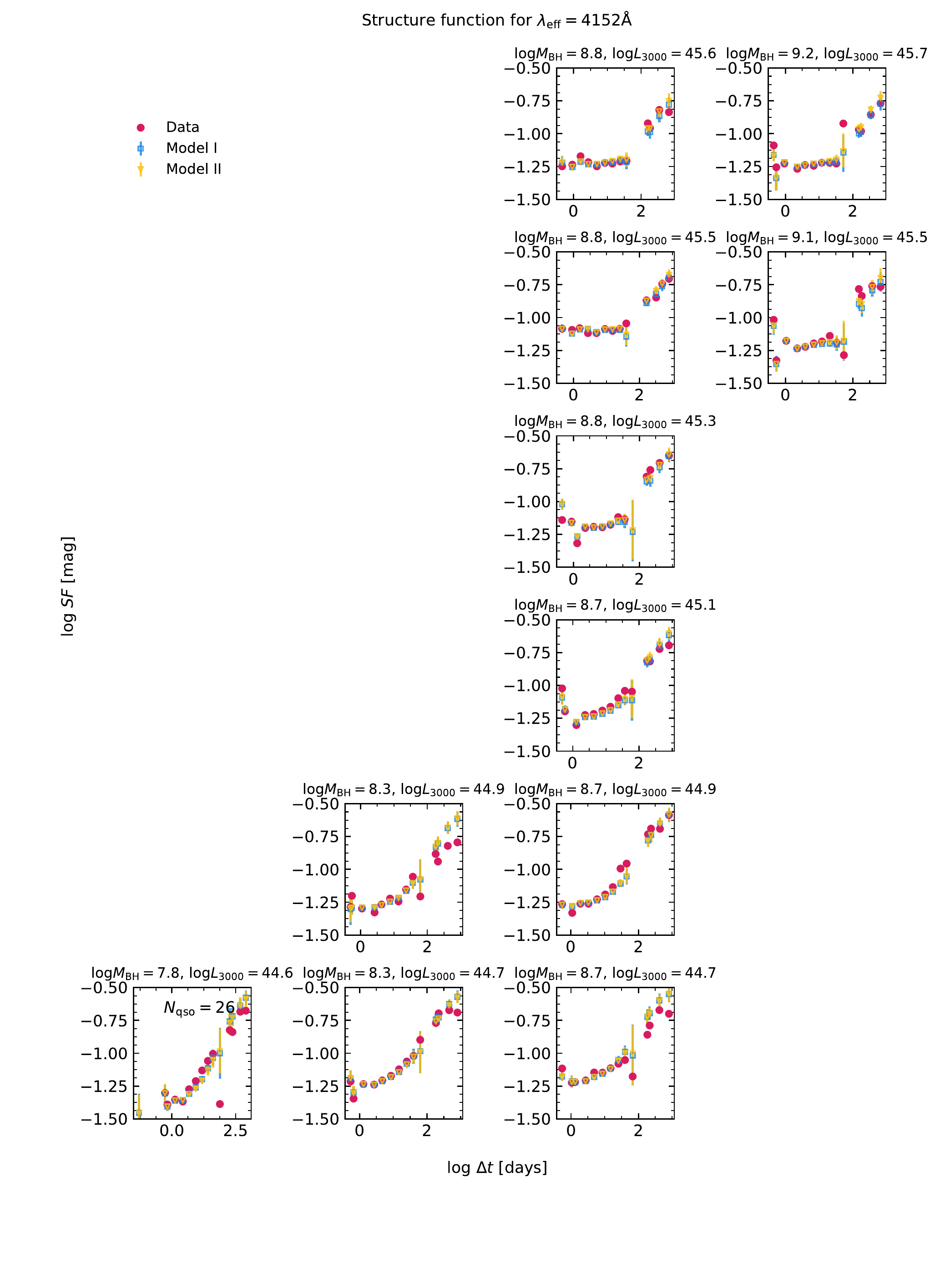}
    \caption{The ensemble structure functions for different [$M_{\mathrm{BH}}$, $L_{3000}$] 
    bins of the $4150\ \mathrm{\AA}$ case. Panels in each column (row) share the same $M_{\mathrm{BH}}$ 
    ($L_{3000}$). 
    \label{fig:sf-D}}
\end{figure*}

\figsetstart
\figsetnum{9}
\figsettitle{The ensemble structure functions (at $\Delta t = 30$ days, $100$ days, and $300$ days) 
as a function of $L_{3000}$ for the four wavelength cases. }
\figsetgrpstart
\figsetgrpnum{9.1}
\figsetgrptitle{The ensemble structure functions at $\Delta t = 30$ days. }
\figsetplot{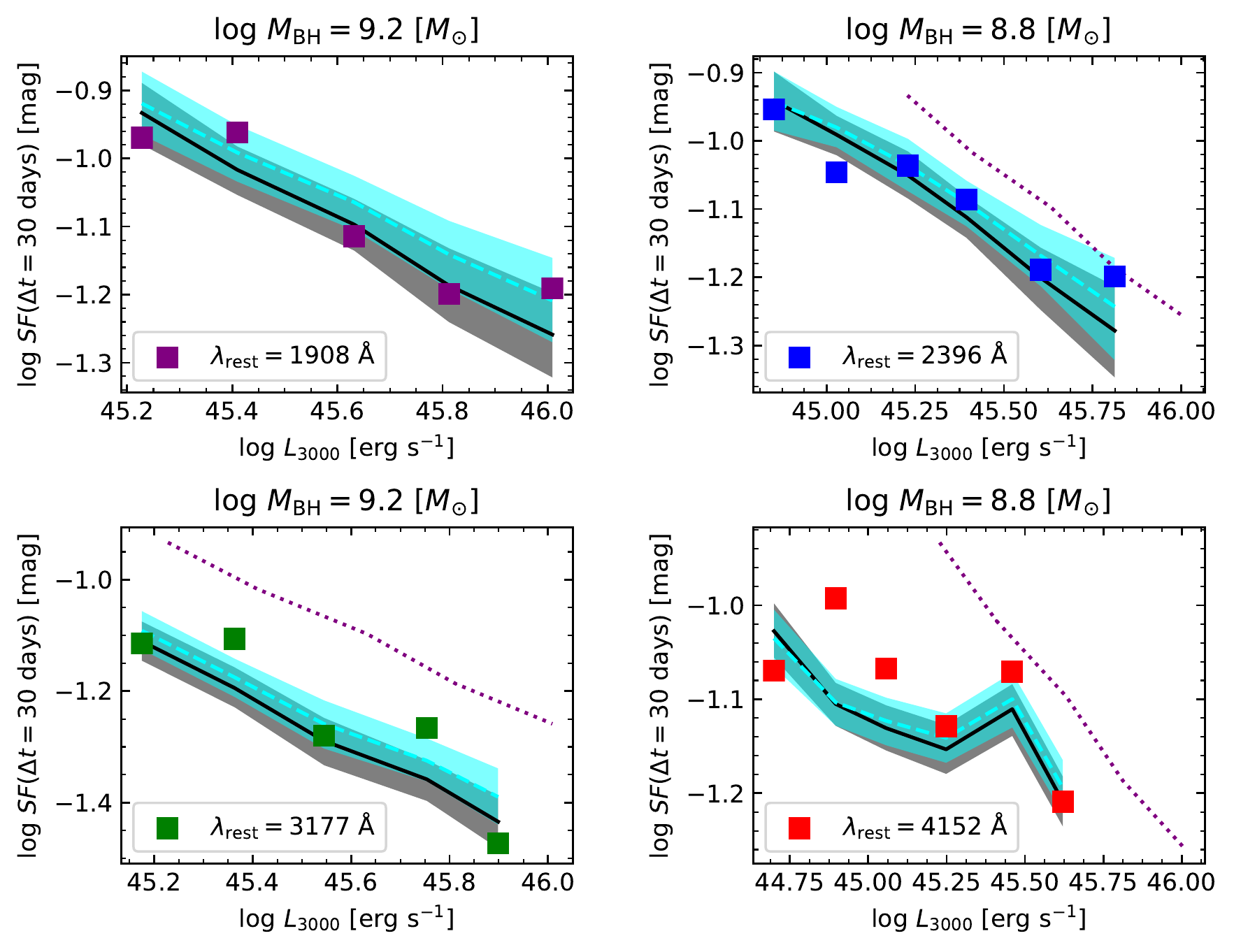}
\figsetgrpnote{he purple, blue, green, and red crosses represent the $1900\ \mathrm{\AA}$, 
$2400\ \mathrm{\AA}$, $3180\ \mathrm{\AA}$, and $4150\ \mathrm{\AA}$ cases, respectively. The black 
solid (cyan dashed) curves and shaded regions correspond to the mean and $1\sigma$ uncertainties of 
the model I (II) results. Note that the observed ensemble structure function for the $1900\ \mathrm{\AA}$ 
case (i.e., the purple dotted curves) is shown in every panel for the purpose of comparison.}
\figsetgrpend
\figsetgrpstart
\figsetgrpnum{9.1}
\figsetgrptitle{The ensemble structure functions at $\Delta t = 100$ days. }
\figsetplot{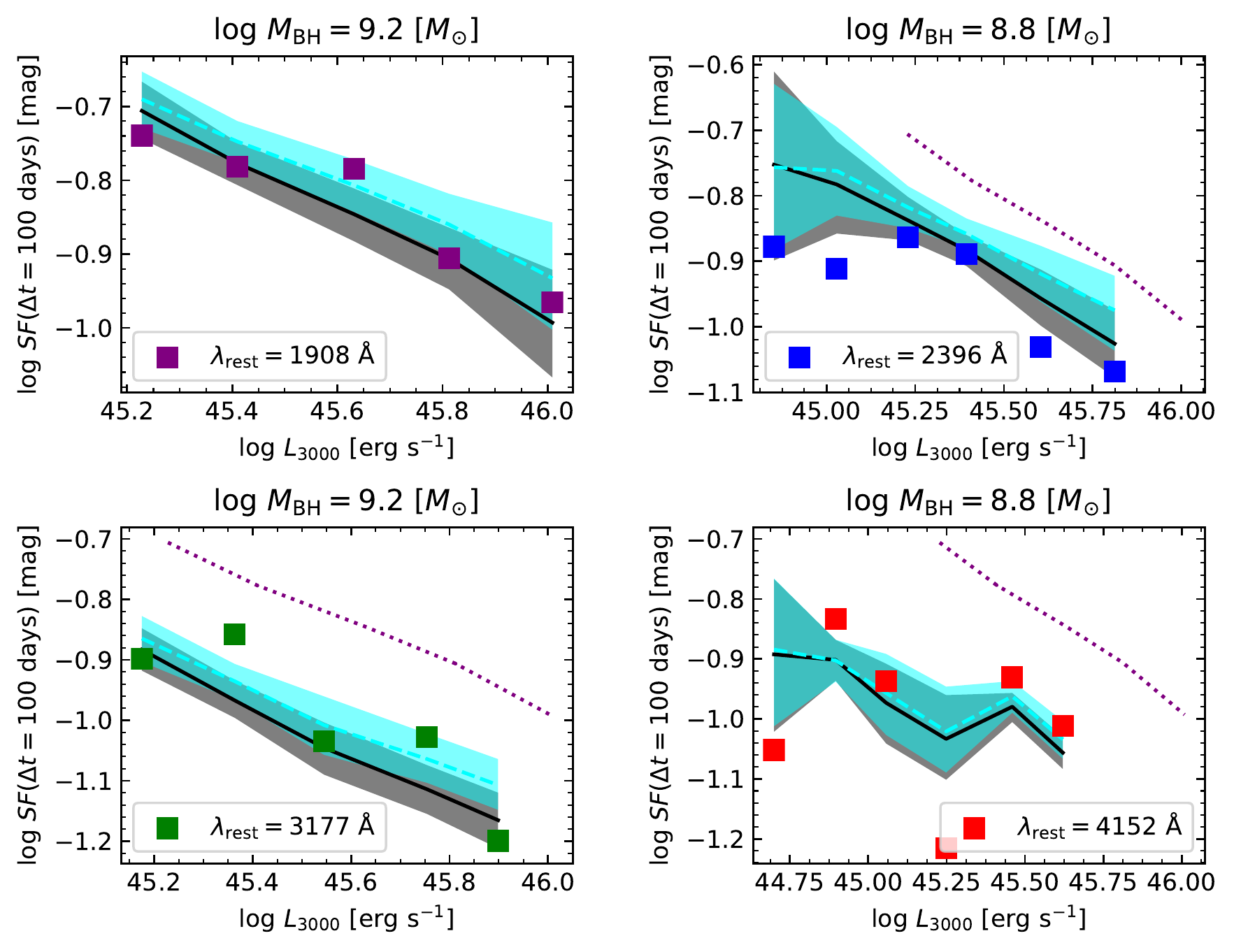}
\figsetgrpnote{he purple, blue, green, and red crosses represent the $1900\ \mathrm{\AA}$, 
$2400\ \mathrm{\AA}$, $3180\ \mathrm{\AA}$, and $4150\ \mathrm{\AA}$ cases, respectively. The black 
solid (cyan dashed) curves and shaded regions correspond to the mean and $1\sigma$ uncertainties of 
the model I (II) results. Note that the observed ensemble structure function for the $1900\ \mathrm{\AA}$ 
case (i.e., the purple dotted curves) is shown in every panel for the purpose of comparison.}
\figsetgrpend
\figsetgrpstart
\figsetgrpnum{9.1}
\figsetgrptitle{The ensemble structure functions at $\Delta t = 300$ days. }
\figsetplot{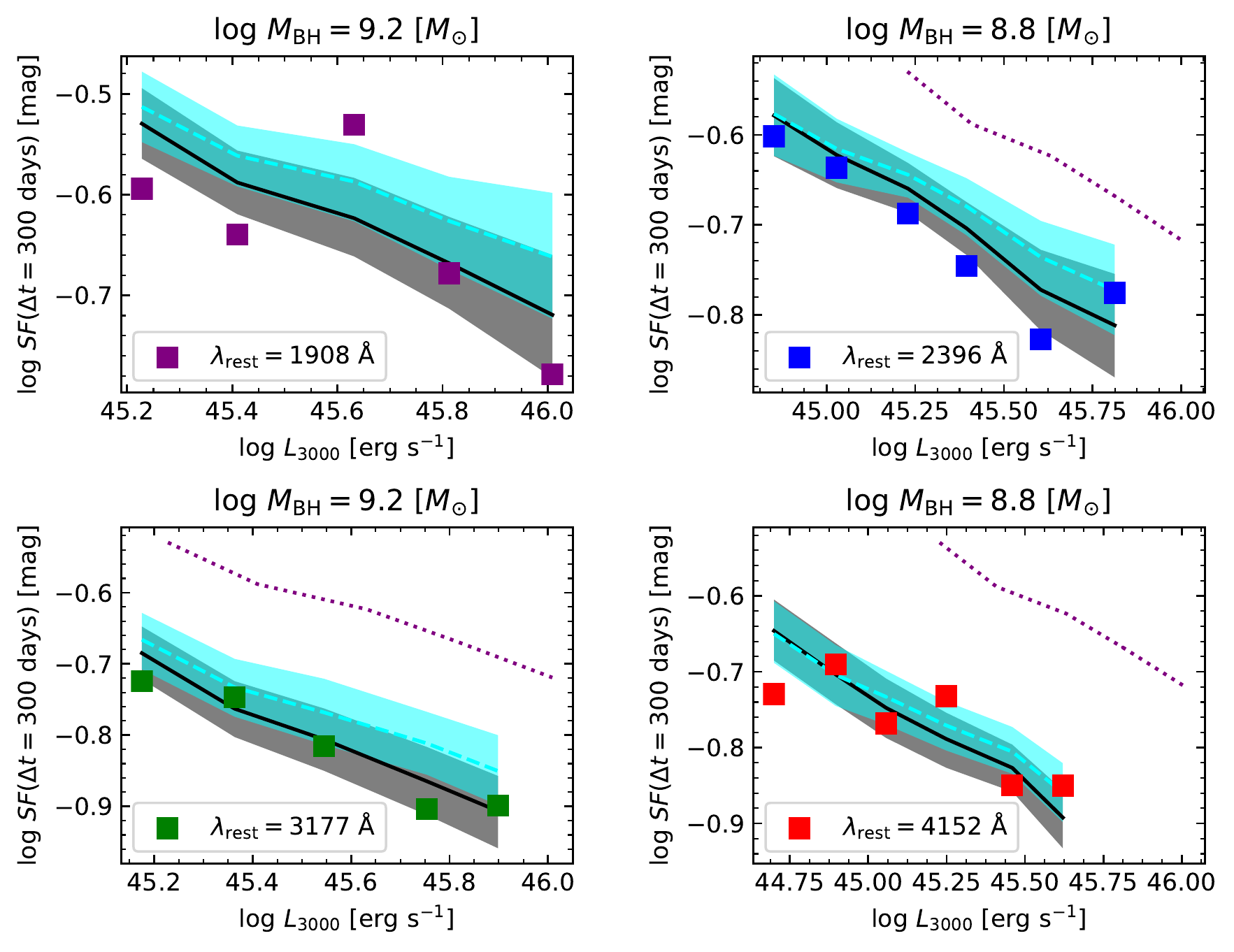}
\figsetgrpnote{The purple, blue, green, and red crosses represent the $1900\ \mathrm{\AA}$, 
$2400\ \mathrm{\AA}$, $3180\ \mathrm{\AA}$, and $4150\ \mathrm{\AA}$ cases, respectively. The black 
solid (cyan dashed) curves and shaded regions correspond to the mean and $1\sigma$ uncertainties of 
the model I (II) results. Note that the observed ensemble structure function for the $1900\ \mathrm{\AA}$ 
case (i.e., the purple dotted curves) is shown in every panel for the purpose of comparison.}
\figsetgrpend
\figsetend

\begin{figure*}
    \epsscale{1.2}
    \plotone{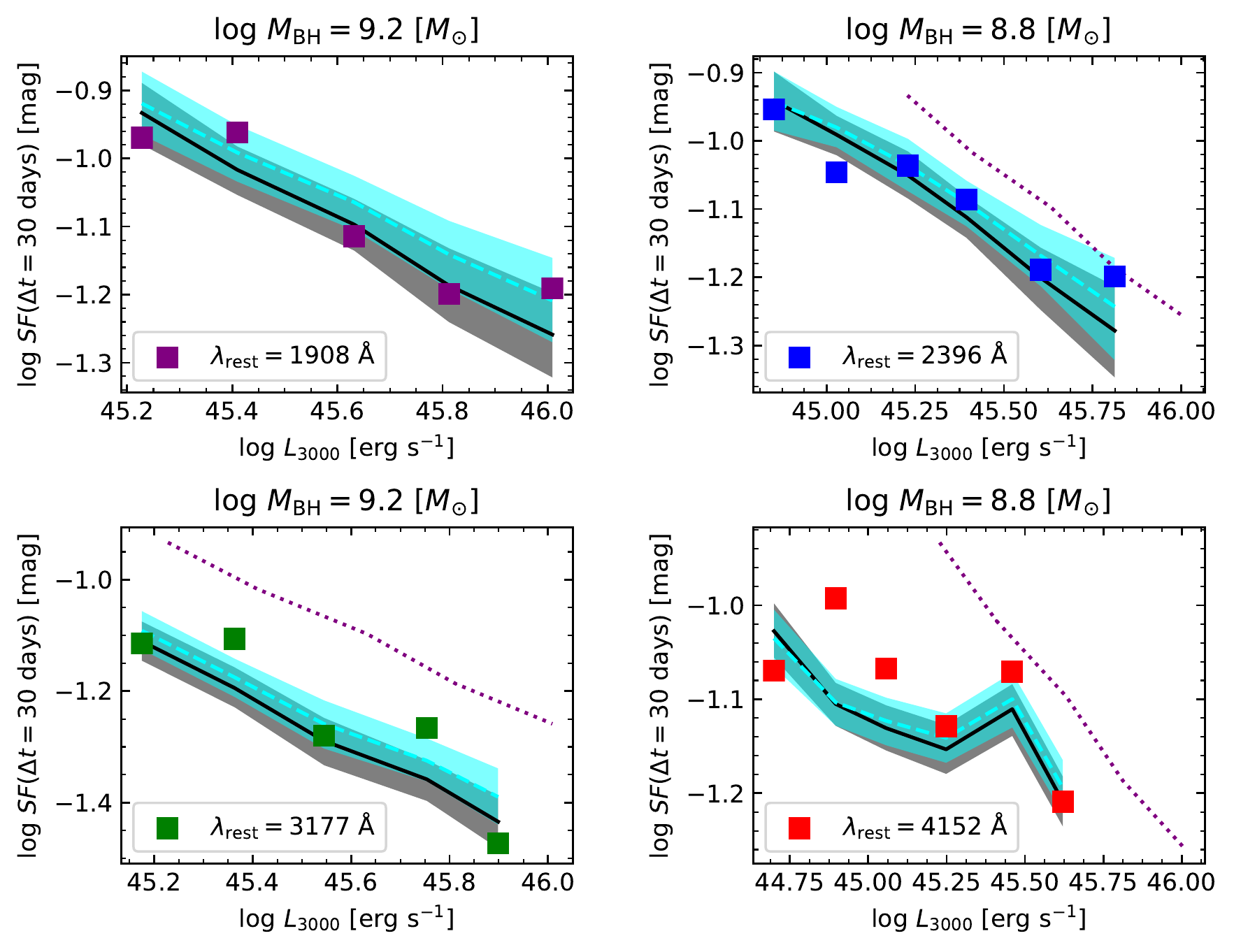}
    \caption{The ensemble structure functions at $\Delta t = 30$ days as a function of $L_{3000}$ 
    for the four wavelength cases. The purple, blue, green, and red squares represent the $1900\ 
    \mathrm{\AA}$, $2400\ \mathrm{\AA}$, $3180\ \mathrm{\AA}$, and $4150\ \mathrm{\AA}$ cases, 
    respectively. The black solid (cyan dashed) curves and shaded regions correspond to the mean 
    and $1\sigma$ uncertainties of the model I (II) results. Note that the model I ensemble structure 
    function for the $1900\ \mathrm{\AA}$ case (i.e., the purple dotted curve) is shown in every panel 
    for the purpose of comparison. The complete figure set (three images) is available in the online 
    journal. 
    \label{fig:sf-pms}}
\end{figure*}

\end{document}